# Resource Provisioning and Scheduling Algorithm for Meeting Cost and Deadline-Constraints of Scientific Workflows in IaaS Clouds

Amit Gajbhiye and Shailendra Singh, *Senior Member, IEEE*

amitgajbhiye03@gmail.com, ssingh@nitttrbpl.ac.in

**Abstract—** Infrastructure as a Service model of cloud computing is a desirable platform for the execution of cost and deadline constrained workflow applications as the elasticity of cloud computing allows large-scale complex scientific workflow applications to scale dynamically according to their deadline requirements. However, scheduling of these multitask workflow jobs in a distributed computing environment is a computationally hard multi-objective combinatorial optimization problem. The critical challenge is to schedule the workflow tasks whilst meeting user quality of service (QoS) requirements and the application's deadline. The existing research work not only fails to address this challenge but also do not incorporate the basic principles of elasticity and heterogeneity of computing resources in cloud environment. In this paper, we propose a resource provisioning and scheduling algorithm to schedule the workflow applications on IaaS clouds to meet application deadline constraints while optimizing the execution cost. The proposed algorithm is based on the nature-inspired population based Intelligent Water Drop (IWD) optimization algorithm. The experimental results in the simulated environment of CloudSim with four real-world workflow applications demonstrates that IWD algorithm schedules workflow tasks with optimized cost within the specified deadlines. Moreover, the IWD algorithm converges fast to near optimal solution.

## 1 INTRODUCTION

WORKFLOWS represented as directed acyclic graphs are used to model variety of scientific and business applications. Cloud computing infrastructure presents an interesting prospect for executing these data and compute intensive workflow applications. Workflows are coarse-grained parallel applications that consist of a series of computational tasks logically connected by data- and control-flow dependencies [1]. These arrangements of computational processes are used in the analysis of large scale scientific applications in the diverse fields of astronomy, earthquake science and bioinformatics [2]. Scientific applications are best modelled by workflows due to their distributed nature. Workflow combines several different computational processes into a single logical process. They describe the logical and sequential data and control dependencies and relationships among the individual computational processes of the application. Many large scale projects like Pegasus [3], ASKALON [4], and GrADS [5] have defined workflows to manage and execute them on the Grid.

Cloud computing is a disruptive technology it has changed the way business are conducted and services are delivered to the organizations and individuals. The workflow execution is also influenced by the infrastructure service delivery of cloud computing. Apart from the utility based computing model of cloud computing which allows the users to pay for what they use there are other advantages of executing workflows in cloud computing. First, the on demand provisioning of resources in the cloud computing allows the workflow management system to provision as many as resource required to execute the workflows. The elasticity of resource provisioning allows the workflow management system to scale up or down as and when required in the cloud environment. Second, for provisioning of resources to schedule the workflow computation tasks grids and clusters uses a batch scheduler which employs job queues to hold workflow tasks until the resources are available to schedule the jobs. As a result, the jobs which require large number of computing resources or have long runtimes faces long, unpredictable computation times. On the contrary, in cloud computing model the users provisions resources once and uses them to execute tasks many times. This provisioning model improves the workflow execution performance as there is less scheduling overhead and less waiting time in queue. Third, the use of virtualization in cloud environments allows the heterogeneous legacy components to work on a single platform. Virtualization also allows the workflows to be bundled up and redeployed on the cloud as per needs.

Even though the cloud environment gives the illusion of infinite computing resources with the elasticity to scale up or down the computing resources dynamically according to the needs of workflow management system it comes with the challenge of efficient utilization of cloud resources to the service provider. One of the most challenging NP-Complete problems [6] that researchers try to solve is the problem of scheduling jobs on the available computing resources optimizing one or more scheduling parameters simultaneously. Workflow scheduling is the problem of allocating computing resources to each task and determining the order of execution to satisfy some performance criterion. For the traditional distributed systems like Grids the most common performance criterion is the minimization of total execution time (makespan) of the workflow [7]. For business and market-oriented clouds there is one more prominent scheduling parameter which is economic cost when it comes to executing workflows in cloud computing environment. From the end users perspective the minimization of both the execution time and execution cost scheduling parameters are preferred. However, in cloud computing the faster resources required to reduce the execution time are more expensive as compared to slower resources and when slower resources are employed to execute the workflows they compromise the makespan of the workflows. Thus there is trade-off between time of execution and cost when it comes to scheduling workflows in clouds. The problem of optimizing time and cost parameters simultaneously belongs to multi-objective optimization problem in cloud computing.

When the workflows are executed in cloud environment there are mainly two stages [8]. First, the resources on which the computation tasks will be deployed will be selected and provisioned on the basis of user defined Quality-Of-Service (QoS) parameters this process is called resource provisioning. Second is the stage of scheduling in which the tasks are actually deployed on the best resources provisioned in the first stage. When the workflows are executed in Girds the workflow management system knows in advance the availability of static resources and their configuration and hence more emphasis is given on the efficient allocation of these resources to the tasks so as to minimize the execution time (makespan). However, in cloud computing users are empowered with the facility to provision recourses according to their requirement and can reduce cost of workflow execution.

In this paper, we aim to answer the pivotal research challenge of minimizing the execution cost of workflows while meeting the user defined Quality-of-Service (QoS) parameter of workflow deadline. To achieve our aim, we have developed a cost minimized deadline constrained heuristic resource provisioning and scheduling algorithm by the application of bio-inspired swarm-based optimization algorithm called Intelligent Water Drop (IWD) algorithm. Most of the previous research work focused on only scheduling of jobs for executing workflow applications in distributed environment. To meet the challenges of resource provisioning and minimizing the workflow execution cost, in this research work we have combined the resource provisioning and scheduling problem as an optimization problem. Our model considers the real time challenges of virtual machine performance variation, heterogeneity and dynamic provisioning of unlimited computing resources.

The IWD algorithm was first proposed by Shah Hosseini [9] and is inspired by the dynamics of river system and the actions and reactions that occur among the water drops and the water drops with the riverbeds. The IWD algorithm has been applied successfully to solve optimization problems of travelling salesman's problem, the n-queen problem and the multiple knapsack problems in [10]. It has also being employed in Vehicle Routing Problem [11], Mitigating Distributed Denial of Service (DDoS) attacks [12], Air Robot Path Planning [13] and many other optimization problems.

The rest of the paper is organized as follows. Section 2 discusses the related work. In Section 3 we formulate the problem of resource provisioning and scheduling and elaborates the application and resource models. Section 4 is a brief description of Intelligent Water Drop algorithm. Section 5 discusses system modeling, schedule generation process and the IWD algorithm for resource provisioning and scheduling. Finally, Section 6 reports the results and analysis of performance evaluation of the algorithm followed by Section 7 which presents conclusion and future work.

## 2 Related Work

Provisioning of resources and scheduling of workflow application in cloud computing is a crucial research challenge and hence it is widely studied by the research community. As scheduling workflows in distributed environment is an NP-complete problem many heuristic and meta-heuristic algorithms have been developed to solve this problem. Workflow scheduling algorithms in high performance distributed computing environment can be mainly categorized as best-effort based and Quality of Service (QoS) constraint based [14]. Best-effort based scheduling algorithms primarily focus on minimizing the workflow execution time without considering any of QoS parameters like execution cost. The prominent algorithms in this category are HEFT [15], Min-Min algorithm [16] and a throughput maximization strategy [17] used by SwinDeW-G [18] and SwinDeW-C [19]. In contrast, the QoS constraint based scheduling algorithms schedules the workflows optimizing the user define QoS constraints. Because of the pay-as-you-go pricing model of cloud computing the main goal of any resource provisioning strategy is to minimize the execution cost of workflow applications while meeting application QoS constraints. Hence one of the important scheduling parameter in QoS based scheduling algorithm is cost of execution of workflow from user perspective. Most of the scheduling algorithms in QoS constraint based scheduling either minimize time under budget constraints or minimize cost under deadline constraints [20]. Our work in this paper focuses on latter of the two QoS constraint based scheduling algorithms. Some of the prominent scheduling algorithms in the QoS constraint based algorithms are back-tracking [21] algorithm, the LOSS and GAIN approach implemented in CoreGrid [22], a genetic algorithm approach [23] and the deadline distribution algorithm [24] implemented in GridBus [25]. Grids have been a major infrastructure provider for executing scientific and business application workflows before the cloud computing framework. A comprehensive survey of workflow scheduling algorithms has been done in [14]. A QoS constraint based dynamic critical path workflow scheduling algorithm has been proposed in [26]. The algorithm works by calculating the critical path in the workflow directed acyclic graph at each step and outperforms the other existing heuristic and meta-heuristic based scheduling strategies especially when resource availability is dynamic. The work in [27] optimizes user preferred multiple QoS parameter based on Ant Colony Optimization. In [23] a budget constrained scheduling algorithm which minimizes execution time is developed for utility Grids.

A vast majority of aforementioned scheduling algorithms in Grid environment are best effort based algorithms and thus concentrates on reducing the application execution time. There are various characteristic of cloud environment like heterogeneity of virtual resources and their different prices, and performance variation of leased cloud resources which separates clouds from grids. Moreover the heavy use of virtualization in cloud environment separates clouds from grid technology [28]. Thus the resource provisioning and scheduling algorithms developed for grids are not suited for cloud environment.

As stated this research work is related to cost-minimization deadline-constrained workflow scheduling developed for IaaS cloud environment. The recent work highly relevant to our work is done in [29]. The authors in this paper have considered virtual machine instances of various size/cost are available and the goal is to finish executing the jobs within their deadlines at the minimum cost. The auto-scaling mechanism in this research work dynamically monitors the workload of the system and progress of jobs and then makes the scheduling decision based on updated information. However they assume that the users have unlimited budget and they try to minimize the cost while maintaining the target level of services. This may not be the scenario with many real time cloud users as their budget may be limited and the schedule generated by this algorithm may exceed their defined budget limits.

The authors in [30] have developed dynamic, Dynamic Provisioning Dynamic Scheduling (DPDS), Workflow-Aware DPDS (WA-DPDS) and static, Static Provisioning Static Scheduling (SPSS) algorithms for resource provisioning and task scheduling for the workflow ensembles in IaaS clouds. As contrast to the work [29] they considered a given budget and try to allocate the virtual machines to execute the workflow within deadline with least possible cost.

In [7] the authors have enhanced their previous algorithm developed for utility Grids to minimize the workflow execution cost while meeting user define deadline. In their work for IaaS clouds they have developed two algorithms named, one phase IaaS Cloud Partial Critical Paths (IC-PCP) and two phase IaaS Cloud Partial Critical Paths with Deadline Distribution (IC-PCPD2). The algorithms consider the unique features of pay-as-you-go pricing model, on-demand resource provisioning and heterogeneity of cloud environments. The algorithm finds the overall real critical path of the workflow beginning from the exit node ending in entry node of the workflow graph and then tries to schedule all the tasks on the critical path to a single instance of computation service which can execute the tasks with minimum cost within the deadline. The algorithm saves the data transfer time as it schedule all the tasks on the critical path on the same computation instance.

Finally, in the research work [31] a Partitioned Balanced Time Scheduling (PBTS) algorithm is proposed which determines the optimal number of computing resources so as to reduce the total cost of workflow execution. The algorithm reduces the number of resources and hence cost of execution by introducing artificial slake time in the execution of a task as long as this does not violate the deadline of workflow. Based on the current running status of VMs and tasks in the system the algorithm dynamically generates resource schedule every charge interval.

## 3 PROBLEM FORMULATION

### 3.1 Application and Resource Models

For the scheduling system model in this research work, the workflow application $W = (T, E)$ is modeled as a directed acyclic graph where $T = \{t_1, t_2, \dots t_n\}$ is the set of $n$ tasks and $E$ is the set of edges representing data and/or control dependencies between the tasks. An edge $e_{ij}$ represented by $(t_i, t_j)$ between two task nodes represents the control and data flow dependencies between tasks. Moreover, the dependency between the task $t_i$ and $t_j$ suggests that $t_j$ can not execute before $t_i$. Task $t_i$ is called parent task and $t_j$ is called child task. A task with no parent is called entry task and a task with no child is called exit task. The workflow $W$ has a deadline $\delta_W$ which represents the latest finish time of the workflow. A sample workflow is depicted in Fig. 1. For our research we follow the resource model as is developed in [9]. Our cloud model for scheduling consists of an IaaS service provider which provides virtual machines (VM) with varying processing capacity. We assume that the service provider provides services similar to Amazon Elastic Compute Cloud (EC2) [32]. It also has a storage service similar to Amazon Elastic Block Store (EBS) [33] which acts as local storage to resources and facilitates the input and output of files as local files. We assume that we are provided with the processing power of VMs in millions of instruction per second (MIPS) from the provider. We use the processing power of VM to calculate the execution time of task on the virtual machine. Furthermore, we assume that the executing task on the resources has exclusive access to the virtual machine instance and the execution of task is non pre-emptive. As is the case with current EC2 service the VMs are charged in integral time units. Any partial utilization of virtual machine bears charge for full time period. For example if the time period is one hour and the virtual machine is employed for 125 minutes the user will be charged for three time periods. A user can lease as many as VMs as is required for the execution of the workflow.

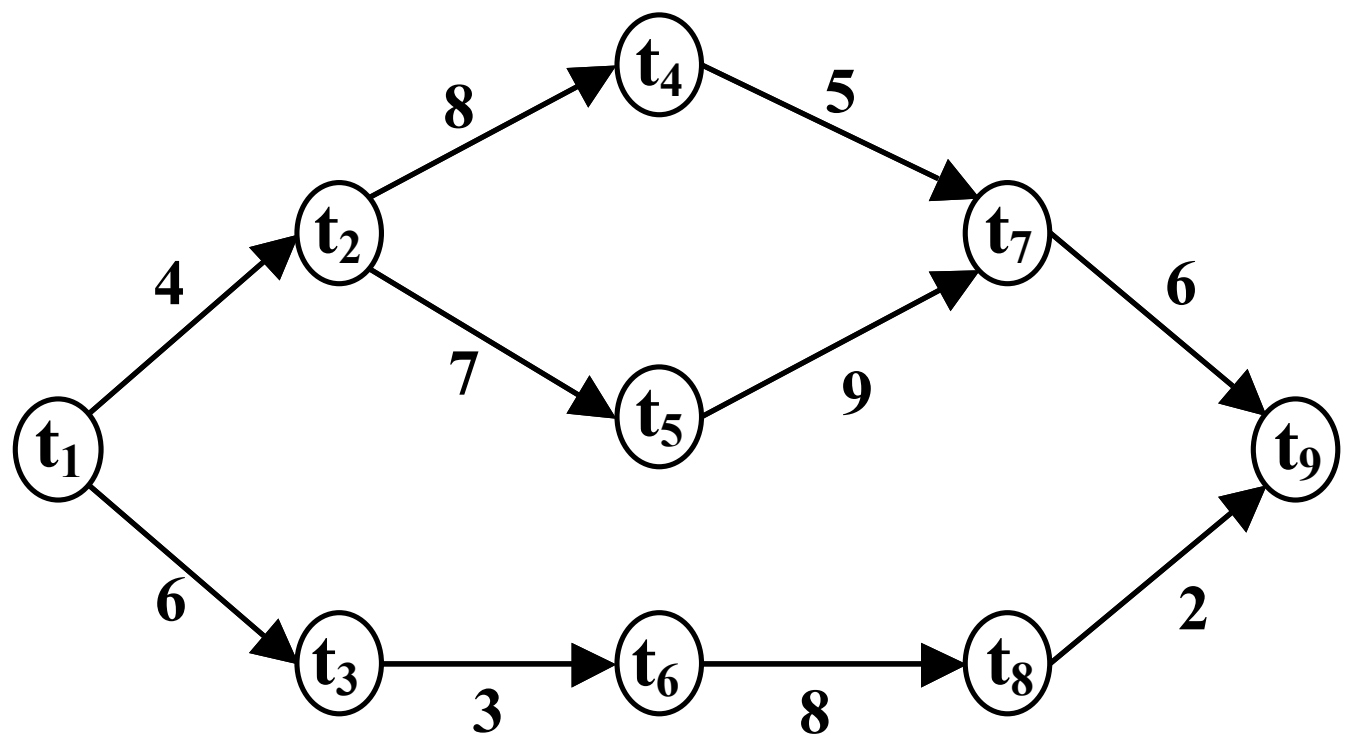

Fig. 1 An example workflow represented as a directed acyclic graph. The nodes of the graph represent tasks of the graph and edges shows control and data dependency between the tasks. The number on the edges shows the data transfer time.

We define $ET_{ti}^{VMj}$ as the execution time task $t_i$ on a VM $VM_j$ and are approximated using the task size $I_{t_i}$ in MIPS. The $ET_{ti}^{VMj}$ is calculated according to Equation (1). Additionally, we assume that all the computation and storage is done in the same datacenter or region so the average bandwidth ($\beta$) between the VM is almost same. The time $TT_{e_{ij}}$ taken by the parent task to transfer $d_{t_i}^{out}$ to the child task is depicted by equation (2). The data transfer time between two tasks executing in the same virtual machine is zero. Finally, the total processing time of a task $PT_{ti}^{VMj}$ is the sum of execution time of task and the time it takes to receive data from all of its parent task and is given by Equation (3) where k is the number of edges in which $t_i$ is a parent task and $s_k$ determines whether the tasks execute on the same virtual machine or not. When the tasks execute on the same virtual machine $s_k$ is 0 and otherwise it is 1.

$$ET_{t_i}^{VM_j} = \frac{I_{t_i}}{\left(P_{VM_j} \times \left(1 - deg_{vm_j}\right)\right)} \tag{1}$$

$$TT_{e_{ij}} = \frac{d_{t_i}^{out}}{\beta} \tag{2}$$

$$PT_{t_i}^{VM_j} = ET_{t_i}^{VM_j} + \left(\sum_1^k TT_{e_{ij}} \times s_k\right) \tag{3}$$

Furthermore, many commercial cloud service providers do not charge for data transfers within the datacenter as a result we also do not consider this cost when calculating the workflow execution cost. In the next section, we formally define the scheduling problem as an optimization problem.

### 3.2 Problem Definition

Our research work focuses on developing a heuristic algorithm for deadline-constrained cost-minimization workflow scheduling for IaaS cloud environment. To model the problem of scheduling as an optimization problem we define a schedule $S = (R, M, TEC, TET)$ where R is the set of resources, M is a mapping from task to resource, TEC is the Total Execution Cost, TET is the Total Execution Time. The resource set $R = \{r_1, r_2, r_3, \dots, r_n\}$ is the set of virtual machines which are available on rent. Each virtual machine has a lease start time $LST_{r_i}$ and a lease end time $LET_{r_i}$. For each workflow task the mapping M is represented by the form $m_{ti}^{rj} = (t_i, r_j, ST_{t_i}, ET_{t_i})$. The mapping represents that the task $t_j$ is allocated to resource $r_j$ with the expected start time of execution $ST_{t_i}$ and completion time $ET_{t_i}$. Equations (4) and (5) shows the total execution cost (TEC) and total execution time (TET).

$$TEC = \sum_{i=1}^{|R|} C_{VM_{r_i}} \times \left\lfloor \frac{(LET_{r_i} - LST_{r_i})}{\tau} \right\rfloor \tag{4}$$

$$TET = \max \left\{ ET_{t_i} : t_i \in T \right\} \tag{5}$$

Thus on the lines of the goal of this research work of developing a heuristic algorithm for deadline-constrained cost-minimization workflow scheduling for IaaS cloud environment we can now formally define the scheduling problem as follows: find a schedule S which minimizes the value of TEC with the value of TET within the workflow deadline $\delta_W$. Mathematically, it is represented as in equation (6).

$$\text{Minimize TEC} \quad \text{subject to TET} \le \delta_W \tag{6}$$

## 4 Intelligent Water Drop Algorithm (IWD)

### 4.1 Principles of IWD

Intelligent Water Drop (IWD) algorithm is nature-inspired swarm intelligence based meta-heuristic algorithm [34]. It was first proposed in [9] as a problem solving algorithm and is employed to solve some artificial and standard versions of Travelling Salesman Problem yielding promising near optimal solutions to the problem. The algorithm is inspired by the process that takes place in the Nature between the flowing water drops of river and the soil of river bed. In IWD algorithm each water drop has two dynamic properties: the velocity with which the water drops move in the water and the amount of soil it carries. The IWDs are assumed to be moving in discrete finite length steps. An IWD begins its journey from source with a finite amount of initial velocity and zero amount of soil content to a destination with some finite velocity and finite amount of soil. When the IWD moves from its current location to the next location its velocity is increased and this increase in velocity is non-linearly proportional to the inverse of the soil between the source and destination locations. As a result IWD travels faster on a path with less soil than a path with more soil. An IWD accumulates soil from the riverbed during its flow in the environment. This amount of soil is non-linearly and inversely proportional to the amount of time the IWD takes to move from its current location to the next location. The time taken by the IWD to move from one location to another location is governed by the laws of liner motion and thus is proportional to the velocity of IWD and is inversely related to the distance between two locations. Furthermore, an IWD selects a path with lower soil on the riverbed compared to the path with higher soil on its bed.

### 4.2 The IWD Algorithm

Based on the above observations the IWD algorithm is developed. The problem to be solved is represented as a graph $G = (N, E)$, where $N$ is the set of nodes in the graph and $E$ is the set of edges between the nodes. Initially the IWDs are placed on the random node of the graph. Each IWD then starts travelling to the nodes of the graph constructing the solution. The iteration end is marked by the completion of solution by each of the IWD. After each iteration, the iteration best solution ($T^{IB}$) is compared with the overall best solution found yet (called total best solution ($T^{TB}$) and the total best solution is updated with the iteration best solution if it is better than total best solution. To allow more number of IWDs to take optimal path the soil on the iteration best solution is reduced commensurate to the goodness of solution. The next iteration then begins with new IWDs and updated soil on the paths from the previous iteration. The algorithm terminates when desired quality solution is found or when the maximum number of iterations reached. The IWD algorithm employs static and dynamic parameters. Static parameters remain same throughout the algorithm execution and dynamic parameters change after every iteration of the algorithm.

## 5 PROPOSED APPROACH

### 5.1 System Model

We have modeled the IWD algorithm to provision resource and schedule the tasks of workflow on the IaaS service model of cloud. As IWD is a constructive method it constructs an optimal solution by the cooperation among the swarm of water drops. For provisioning of resources for executing workflows in a cost and time constrained manner the system is available with a number of VMs and from among them we have to choose a set of VMs which can execute the tasks with least cost while meeting workflow's deadlines. The tasks of a workflow are modeled with the IWDs and they flow over a fully connected graph of VMs called VM construction graph give by $G = (N, E)$ progressively constructing the solution. A sample VM construction graph is shown in Fig. 2.

### 5.2 Schedule Generation

For our research we have modeled the water drops as tasks of the workflow in the cloud environment. The available set of virtual machines is assumed to form a fully connected graph $G = (N, E)$ called Virtual Machine (VM) construction graph. Initially the resource set to rent $R$ and the set of task to resource mapping $M$ is empty and the total execution time $TET$ and total execution cost $TEC$ of workflow are set to zero.

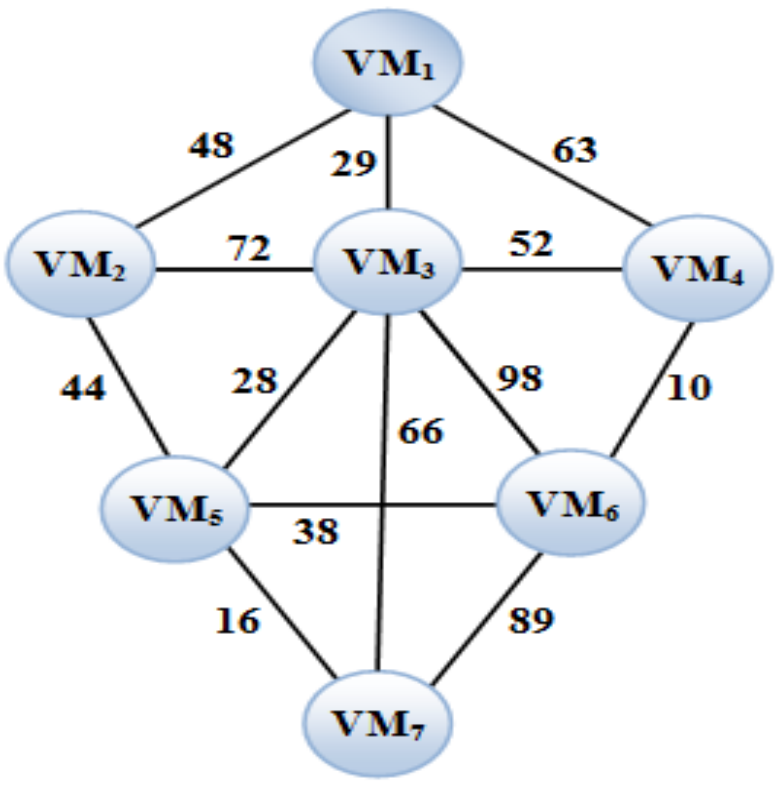

Fig.2 Sample VM Constructions Graph. The numerical value on edges show the local soil of the path

The algorithm generates a feasible solution schedule in Solution Construction Phase of IWD algorithm. In an iteration of the solution construction phase of our algorithm we generate a schedule and calculate the cost of executing the schedule and the time of execution of workflow on the VMs visited by the task. The execution cost and the execution time for the generated schedule is stored for comparison with the schedules generated in the subsequent iterations.

In our model of resource provisioning, for each iteration of the IWD algorithm each IWD visits a fixed predefined number of VMs from the whole pool of available VMs. This fixed number reduces the time required in provisioning the resources. Each IWD (task) has a list of visited nodes and it also notes its cost of execution and the time it takes to execute on the visited node (VM). Thus each IWD visits the fixed number of VMs noting the execution cost and time it takes to execute on that virtual machine.

The execution time is calculated according to the formula shown in Equation (1). Iteration ends when all the IWDs have visited their fixed number of nodes. As in our model we know the amount of output data $d_{ti}^{out}$ produced by each task $t_i$ and we have uniform bandwidth available among the VMs so we also calculate the data transfer time $TT_{e_{ij}}$ according to Equation (2). For each task we calculate the data transfer time $\left(TT_{e_{ij}}\right)$ between a parent task $t_i$ and child task $t_j$. With the help of $TT_{e_{ij}}$ we can now calculate the total processing time $PT_{ti}^{VMj}$ task $t_i$ on virtual machine $VM_j$ according to Equation (3). The cost of executing the task on the resource $r_j$ is calculated according to Equation (7).

$$C_{t_i}^{r_j} = ET_{t_i}^{VM_j} \times C_{VM_{r_j}} \quad (7)$$

At the end of the iteration for each task we have the data structures shown in Fig. 3 which contains the task to resource mapping, execution time and the cost of executing task on the particular resource (VM) visited. Thus at the end of each solution construction phase we have a schedule. For this schedule we sum the minimum cost of execution for each task $\left(\min\left\{execCost_{t_i}^{r_j}\right\}\right)$ and their corresponding time of execution $\left(execTime_{t_i}^{r_j}\right)$.

If the total time of execution is within the user specified deadline ($\delta_W$) and the cost obtained for this iteration ($Cost(T^{Itr})$) is less than iteration best solution cost ($Cost(T^{IB})$) we update the iteration best solution schedule ($T^{IB}$) with this iteration solution ($T^{Iter}$) and update the soil value for all edges included in the ($T^{IB}$) otherwise we discard the schedule and

$taskToResourceMap_{t_1}^{r_j} =$

| | | | | | | | | | |
|---|---|---|---|---|---|---|---|---|---|
| $r_6$ | $r_4$ | $r_3$ | $r_1$ | $r_8$ | $r_2$ | $r_5$ | $r_7$ | $r_9$ | $r_{11}$ |

(a)

$execTime_{t_1}^{r_j} =$

| R | $r_1$ | $r_4$ | $r_3$ | $r_6$ | $r_8$ | $r_2$ | $r_5$ | $r_7$ | $r_9$ | $r_{11}$ |
|---|---|---|---|---|---|---|---|---|---|---|
| ET | 5 | 6 | 2 | 9 | 4 | 7 | 6 | 4 | 3 | 8 |

(b)

$execCost_{t_1}^{r_j} =$

| R | $r_1$ | $r_4$ | $r_3$ | $r_6$ | $r_8$ | $r_2$ | $r_5$ | $r_7$ | $r_9$ | $r_{11}$ |
|---|---|---|---|---|---|---|---|---|---|---|
| EC | 5 | 6 | 2 | 9 | 4 | 7 | 6 | 4 | 3 | 8 |

(c)

$transferTime =$

| | t1 | t2 | t3 | t4 | t5 | t6 | t7 | t8 | t9 |
|---|---|---|---|---|---|---|---|---|---|
| t1 | 0 | 4 | 6 | 0 | 0 | 0 | 0 | 0 | 0 |
| t2 | 0 | 0 | 0 | 8 | 7 | 0 | 0 | 0 | 0 |
| t3 | 0 | 0 | 0 | 0 | 0 | 3 | 0 | 0 | 0 |
| t4 | 0 | 0 | 0 | 0 | 0 | 0 | 5 | 0 | 0 |
| t5 | 0 | 0 | 0 | 0 | 0 | 0 | 9 | 0 | 0 |
| t6 | 0 | 0 | 0 | 0 | 0 | 0 | 0 | 8 | 0 |
| t7 | 0 | 0 | 0 | 0 | 0 | 0 | 0 | 0 | 6 |
| t8 | 0 | 0 | 0 | 0 | 0 | 0 | 0 | 0 | 2 |
| t9 | 0 | 0 | 0 | 0 | 0 | 0 | 0 | 0 | 0 |

Fig. 3. Sample data structure after an iteration for $t_1$. (a) Task to resource mapping. (b) Execution time (c) Cost of execution. (d) Transfer Time Matrix

proceed to next iteration. Equation (8) shows the mathematical expression for the calculation of minimum cost of schedule and condition for updating total best solution with iteration best solution.

$$\sum_{i=1,j=1}^{i=N,j=|R|} \min\left\{execCost_{t_i}^{r_j}\right\}$$

$$\text{Subject to} \sum_{i=1,j=1}^{i=N,j=|R|} execTime_{t_i}^{r_j} < \delta_W$$

$$\text{and} \sum_{i=1,j=1}^{i=N,j=|R|} \min\left\{execCost_{t_i}^{r_j}\right\} < Cost(T^{IB}) \quad (8)$$

The solution construction phase is repeated for $Max_{Itr}$ number of times and finally we update the total best solution ($T^{TB}$) with the most optimal iteration best solution($T^{IB}$). For this near optimal solution we now have the task to resource mapping $taskToResourceMap_{t_j}^{r_j}$, i.e. the first two first two component of our mapping M, we now calculate the other two values of ($ST_{t_i}$) the start time and $\left(ET_{t_i}\right)$ the end time of the task for this resource mapping.

There are two scenarios when calculating $ST_{t_i}$. First, the task has no parent in this case the task can start running as soon as the resource which it was mapped to is available. In the second scenario, the task has one or more parents. In this case the task can start executing as soon as the longest executing parent tasks completes and transfers its data to the waiting child task.

The end time of the task $ET_{t_i}$ is calculated with the help of start time $ST_{t_i}$ and the total processing time of the task. The total processing time of the task is calculated by the summing the execution time of the task taken from Fig. 3(a) data structure and data transfer times from all its parent tasks taken from data structure Fig. 3(d).

At this stage of the algorithm we have calculated all the elements of the mapping M, $\left(t_i,\ r_j,\ ST_{t_i},\ ET_{t_{i_i}}\right)$. In our model we also calculate the time when the VM should be started that is VM lease start time $LST_{r_i}$ and when it should stop processing the tasks that is the lease end time $LET_{r_i}$.

For $LST_{r_i}$, if the resource is already in the list R then it already has a set start time so $LST_{r_i}$ need not to be updated. However, if the resource is not in R and $t_i$ is the first task which is to be scheduled on this resource then $LST_{r_i}$ is set to the VM boot time. This way we also account for the time the VM needs to get ready to execute the task which is considerably high in many real time scenarios.

For the lease end time $LET_{r_i}$, this is the time when the last task ($t_i$) scheduled to execute on it finishes executing. It is calculated by the addition of the time it takes to processes the last task ($t_i$) and the $LST_{r_i}$.

At the end of the algorithm we have all the resources in R which needs to be leased and the mapping of the resources for each task $t_i$ that signifies which task needs to be allocated to which virtual machine and the corresponding start time and end time of virtual machines. With this information, we finally calculate the execution cost TEC and time TET associated to the total best solution $T^{TB}$. The algorithm has now constructed all the elements $(R, M, TEC, TET)$ of the near optimal schedule and it returns this $T^{TB}$ solution schedule.

### 5.3 IWD Algorithm For Resource Provisioning And Scheduling

In this section, we discuss how the IWDs move from one node to another node on the VM construction graph. In our research work this flow of IWDs corresponds to how the task chooses the next VM to visit. We divide the IWD algorithm in to four main phases of initialization, solution construction, reinforcement, and termination. The Algorithm 1 shows the pseudo code for modeling of IWD algorithm for resource provisioning and scheduling of workflow tasks.

The input to the algorithm is a fully connected graph of VMs called VM construction graph, $G = (N, E)$ where the $N = \{r_j | i \in 1,2 \ldots N\}$ is the set of nodes representing resources associated to VMs and $E = \{(i,j) | (i,j) \in N \times N, i \neq j, i, j \in 1,2 \ldots N\}$ is the set of edges between the VMs. The output or solution of the algorithm is a total best solution schedule $T^{TB}$ in the form of $(R, M, TEC, TET)$.

#### 5.3.1 Initialization Phase

The initialization phase of our IWD algorithm is used to initialize the set of static and dynamic parameters of the algorithm. The IWDs are then spread randomly over the VM construction graph.

**Static Parameters:** The static parameters are initialised once during the whole execution of algorithm and they do not change thereafter.

ALGORITHM 1
SCHEDULING HEURISTIC USING IWD ALGORITHM

$Input$: $VM\ Construction\ Graph$
$Output$: $Total\ Best\ Solution\ Schedule\ (T^{TB})$

```
Initialize the static parameters
while number of iteration < Max_Itr do
    Initialize the dynamic parameters
    Spread Iwd number of IWDs randomly on the VM construction graph.
    Update the list V^k_visited to include source node
        while VMs visited < VM_toVisit
            for each Iwd k do
                Update taskToResourceMap^{r_j}_{t_i} to add current visited node r_j
                Calculate execTime^{r_j}_{t_i}
                Calculate execCost^{r_j}_{t_i}
                Select next node according to p^k_i(j)
                Update V^k_visited with next node
                Move drop k from current node to next selected next node
                Update the parameters
                    (a) vel^k(t + 1)
                    (b) soil^k
                    (c) soil(i, j)
            End for
        End while
    Calculate Cost (T^Iter) = Σ_{i=1,j=1}^{i=N,j=|R|} min{execCost^{r_j}_{t_i}}
    if Σ_{i=1,j=1}^{i=N,j=|R|} execTime^{r_j}_{t_i} < δ_W and Cost (T^Iter) < Cost (T^IB)
        Update T^IB = T^Iter
        Update the soil value of all edges included in the T^Iter
End while
Update T^TB = T^IB
for each task t_i calculate ST_{t_i} and ET_{t_i}
for each resource r_j ∈ R calculate LST_{r_i} and LET_{r_i}
return (T^TB)
```

They are:

- **Iwd** : is the number of agents which collaborate to build the solution. In our case this parameter is initialised with the number of tasks of the workflow.
- **Velocity Updating Parameter** $(a_v, b_v, c_v)$: The set of parameters used to control the velocity update function, as depicted in Equation (12).
- **Soil Updating Parameter** $(a_s, b_s, c_s)$: The set of parameters used to control the soil update function of equation (15).
- $\mathbf{Max_{Itr}}$: The parameter depicts the maximum number of iteration the algorithms goes before termination.
- $\mathbf{Init_{Soil}}$: The initial value of local soil.
- **VM_to_Visit**: A user defined fixed value which decides the number of VMs to be visited in the construction phase of the algorithm.

**Dynamic Parameter:** The dynamic parameters are initialized at the beginning of the algorithm and are update during the processes of finding the solution. They reinstate to their initial value after every iteration. They are:

- $\mathbf{V^k_{visited}}$: The node list visited by the water drop k.
- $\mathbf{Init_Vel^k}$: The initial velocity of water drop k.
- $\mathbf{Soil^k}$: The initial soil of water drop k.

### 5.3.2 Solution Construction Phase

This is the most crucial phase of the IWD algorithm which aims to construct the solution schedule of our resource provisioning algorithm. The schedule of provisioning resources explained in the previous section is generated in this phase of algorithm. The nodes are successively added to the resource set R for each water drop as it visits the nodes. The cost of execution and the estimated time of execution are also stored for each visited VM by the task. At each step of the solution construction phase a partial solution schedule is built upon by adding a new node (VM) to the existing partial solution schedule. The construction phase ends when $VM_{to_{Visit}}$ is reached. The construction phase consists of following steps elaborated in the following sections.

#### 5.3.2.1 Edge Selection Mechanism

This mechanism decides the edge to be taken to visit the next VM. The probability of a water drop k taking an edge $e(i,j)$, where $e \in E$ to visit the node j from the current node i is determined by the probability, $p_i^k(j)$ and is calculated as shown in equation (9).

$$p_i^k(j) = \frac{f(soil(i,j))}{\sum_{\forall l \notin V^k_{visited}} f(soil(i,l))} \quad (9)$$

Where $f(soil(i,j))$ is define as in equation (10).

$$f(soil(i,j)) = \frac{1}{\varepsilon + g(soil(i,j))} \quad (10)$$

Where $\varepsilon$ is a small positive number added to the $g(soil(i,j))$ to prevent a possible division by zero.

$$g(soil(i,j)) = \begin{cases} soil(i,j), & \text{if } \min_{\forall l \notin V^k_{visited}} (soil(i,l)) \geq 0 \\ soil(i,j) - \min_{\forall l \notin V^k_{visited}} & \text{otherwise} \end{cases} \quad (11)$$

Where $soil(i,j)$ refers to the amount of soil within the local path between VMs i and j.

#### 5.3.2.2 Velocity and Local Soil Update

When a water drop k moves from node i to j its velocity at the time $(t+1)$ is denoted by $vel^k(t+1)$ is calculated by the Equation (12).

$$vel^k(t+1) = vel^k + \frac{a_v}{b_v + c_v \times soil(i,j)} \quad (12)$$

Where $a_v, b_v, c_v$ are the static parameters used to represent the non linear relationship between the velocity of water drop k $(vel^k)$ to the inverse of the amount of soil of the local path $(soil(i,j)$. As a water drop moves from node i to j the soil it acquires from the riverbed depicted by $soil^k$ and the soil of the local path are updated using the Equations (13) and (14) respectively.

$$soil^k = soil^k + \Delta soil(i,j) \quad (13)$$

$$soil(i,j) = (1 - \rho_n) \times soil(i,j) - \rho_n \times \Delta soil \quad (14)$$

Where $\rho_n$ is a small positive constant which lies between zero and one and $\Delta soil(i,j)$ is the amount of soil carried by the water drop k from the local path between node i and j. Further $\Delta soil(i,j)$ is non linearly proportional to the inverse of $(vel^k)$ of water drop k and is calculated as formulated in Equation (15).

$$\Delta soil(i,j) = \frac{a_s}{b_s + c_s \times time(i,j: vel^k(t+1))} \quad (15)$$

Where $a_s, b_s, c_s$ are the static parameters used to represent the non linear relationship between $\Delta soil(i,j)$ and inverse of the velocity of water drop k $(vel^k)$. The function $time(i,j: vel^k(t+1))$ denotes the time required by the water drop k to move from node i to node j at time $(t+1)$. It is defined as in Equation (16).

$$time(i,j: vel^k(t+1)) = \frac{HUD(i,j)}{vel^k(t+1)} \quad (16)$$

Where $\mathrm{HUD}(i,j)$ is the heuristic undesirability which measures the undesirability of a water drop to move from node i to node j. For our research work $\mathrm{HUD}(i,j)$ is defined as Equation (17).

$$\mathrm{HUD}(i,j) = \frac{ET_{t_i}^{VM_j}}{\delta_W} \tag{17}$$

The processes of selecting the node to be visited and updating the velocity and local soil are repeated for $VM_{to_{Visit}}$ times.

### 5.3.3 Reinforcement Phase

This phase of the algorithm guides the subsequent iterations to follow the most desired path towards the near optimal solution schedule. For the first iteration of our algorithm we update the iteration best solution $(T^{IB})$ with first iteration solution $(T^{Iter})$. For the subsequent iterations we compare the cost of execution of schedule if it is less than the current iteration best solution cost we update the iteration best solution $(T^{IB})$ with this iteration solution $(T^{Iter})$.

$$T^{IB} = \begin{cases} T^{Iter}, & \text{if } C(T^{Iter}) < C(T^{IB}) \\ & \text{and } \sum_{\substack{1 \le i \le N \\ 1 < j < |R|}} execTime_{t_i}^{r_j} < \delta_W \\ T^{IB}, & \text{otherwise} \end{cases} \tag{18}$$

To reinforce the IWDs in the subsequent iterations to follow the path of $T^{IB}$ generating IWDs the soil of all the edged of the current best solution $(T^{Iter})$ is updated with the Equation (19).

$$soil(i,j) = (1 + \rho_{IWD}) \times soil(i,j) - \rho_{IWD} \times soil_{Iter}^{k} \times \frac{1}{q(T^{Iter})} \tag{19}$$

Where $\rho_{IWD}$ is a positive constant. Finally we update the total best solution $(T^{TB})$ with the near optimal solution iteration best $(T^{IB})$ found after $Max_{Itr}$ number of iterations.

### 5.3.4 Termination Phase

This phase marks the completion of the algorithm. The construction and reinforcement phase are repeated until the termination condition is met. We mark the termination of our algorithm when the maximum number of maximum iteration $(Max_{Itr})$ is reached. In this phase we also calculate the other elements of our schedule and return the total best solution $(T^{TB})$.

## 6 Performance Evaluation

In this section, we describe the experiments we have conducted to evaluate the performance of the proposed IWD algorithm. We used CloudSim framework [35] to simulate the IaaS cloud environment. The workflows applications [36] used in the evaluation are Montage (generation of sky mosaics), SIPHT (bioinformatics), LIGO (detection of gravitational wave) and CyberShake (earthquake risk characterization). The algorithms IC-PCP algorithm proposed by Abrishami et al. [7], SCS algorithm proposed by Mao and Humphrey [29] and PSO algorithm proposed by Rodriguez and Buyya [8] are used to compare the performance of IWD algorithm.

The simulation testbed models an IaaS service provider with a datacenter and with virtual machines sizes ranging from small, medium, large and very large. Table 1 depicts the VM configuration used in the simulation and it is based on the current Amazon EC2 offerings. The performance variation of the cloud infrastructure is modeled on the study of Schad et al. [37] . The performance of VM is reduced by at most 24 percent on a normal distribution with mean 12 percent and standard deviation of 10 percent.

TABLE 1
VM CONFIGURATIONS USED IN THE EXPERIMENT

| Type | Memory (GB) | Core Speed (ECU) | Cores | Cost ($) |
|---|---|---|---|---|
| m1.small | 1.7 | 1 | 1 | 0.06 |
| m1.medium | 3.75 | 2 | 1 | 0.12 |
| m1.large | 7.5 | 2 | 2 | 0.24 |
| m1.xlarge | 15 | 2 | 4 | 0.48 |
| m3.xlarge | 15 | 3.25 | 4 | 0.50 |
| m3.xxlarge | 30 | 3.25 | 8 | 1.00 |

The VM time period for billing is chosen to be 1 hour and boot time for a virtual machine is taken as 97 seconds according to the results of [38].

To calculate the deadline of a workflow we followed strategy illustrated in [8]. The tasks of the workflow were first scheduled on a single VM of cheapest time which gives the slowest runtime of the workflow. The fastest execution time for the workflow was calculated similarly but by scheduling workflow tasks on the VM with highest configuration. The deadline interval is defined as the difference of these two deadline divided by five. The first deadline is calculated by adding the deadline interval to the fastest deadline and for second deadline is obtained by adding two deadline intervals and so on. The analysis of algorithm is done from a stricter deadline to more relaxed deadlines.

TABLE 2
STATIC AND DYNAMIC PARAMETERS USED IN IWD ALGORITHM

| Description | Parameters | Values |
|---|---|---|
| Static Parameters | $Iwd$ | Number of Workflow Tasks |
| | $a_v, b_v, c_v$ | 1000, 0.01, 1 |
| | $a_s, b_s, c_s$ | 1000, 0.01, 1 |
| | $Max_{Itr}$ | 20 |
| | $Init_{Soil}$ | 100 |
| | $VM_to_Visit$ | 10 |
| Dynamic Parameters | $V_{visited}^{k}$ | Empty |
| | $Init_Vel^{k}$ | 4 |
| | $Soil^{k}$ | 0 |

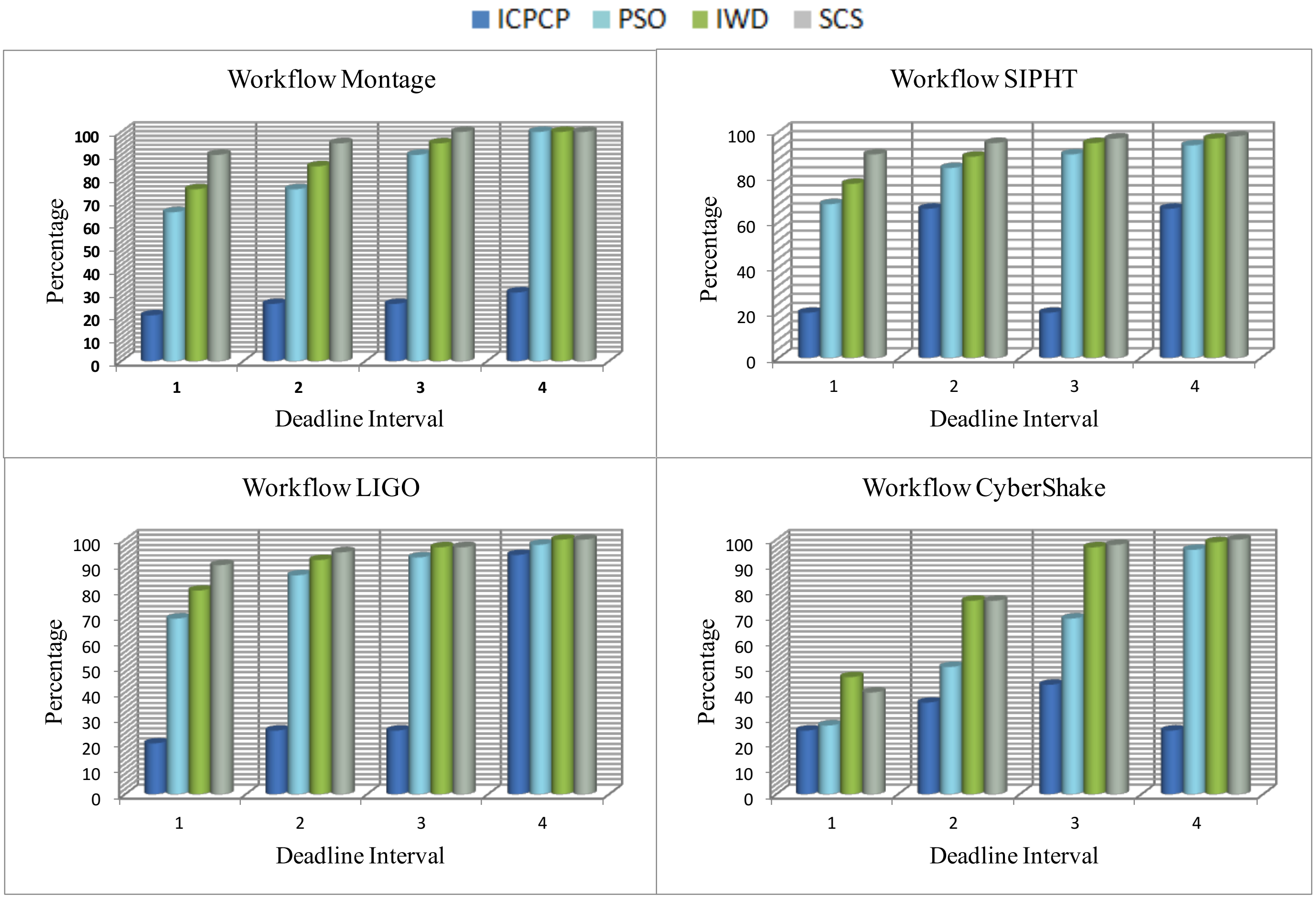

Fig. 4. Percentage of Deadline Met by each workflow in every deadline interval.

**6.1 Results and Analysis**

**6.1.1 Deadline Constrained Evaluation**

Fig. 4 shows the percentage of deadline met by each individual workflow application in each of the four deadlines. For montage workflow ICPCP on an average meets the deadline only 25 percent of times and the highest being 30 percent in the fourth deadline interval. The PSO and IWD algorithms improve their performance over the deadline intervals by meeting all the deadlines in the 4th deadline interval. When compared in individual deadline interval IWD algorithm performed better as compared to PSO and ICPCP by achieving a higher percentage of deadlines meets. In all the IWD and SCS are the best performing algorithms in all the deadline interval and they meet over 95 percent of deadlines in the last two deadline intervals.

For SIPHT workflow, all the algorithms follows the same trend as for the Montage workflow with IWD and SCS being the best performing algorithms by meeting over 95 percent of deadline.

The results for LIGO workflow depicts the same results for first three interval for ICPCP algorithm meeting at most 25 percent of deadline but the algorithm achieves its best performance by meeting 94 percent of deadlines in the 4th deadline interval. The IWD algorithm performs best in the 3rd (97 percent) and 4th (100 percent) interval by meeting deadlines same as SCS algorithm. The PSO and ICPCP algorithms are out performed by IWD and SCS algorithm in all four deadline Intervals.

For the CyberShake workflow, ICPCP algorithm performs better by meeting 43 percent of deadlines in the 3rd interval. The IWD algorithm has outperformed the other three algorithms by meeting the highest of 46 percent of deadlines in the first deadline interval. Our algorithm achieved almost the same performance as that of SCS algorithm for the deadline interval 2, 3 and 4 and has outperformed ICPCP and PSO algorithms.

In essence, the IWD and SCS algorithm has outperformed the ICPCP and PSO algorithm in terms of meeting the deadlines of workflow for all the workflows and in all deadline intervals. The performance of IWD algorithm is very close to that of SCS algorithm and both the algorithm achieve almost 100 percent performance when it comes to meeting the deadlines of workflow applications.

### 6.1.2 Makespan Evaluation

Fig. 5 presents the average makespan values obtained for each workflow in the every deadline interval. For each workflow we have executed each algorithm 20 times and have averaged the makespan. The dotted line in each of the interval shows the corresponding deadline in the interval. For Montage, SIPHT and LIGO the ICPCP algorithm produces the schedules with an average makespan greater than the deadline of workflow for all the intervals and thus fails to meet the deadline. For LIGO, in the deadline interval 1 the PSO algorithm produces some schedules which exceeds the deadline of workflow but the performance of the algorithm improves considerably in 3rd and 4th deadline interval and the schedule produced are well within the specified deadline of the workflow.

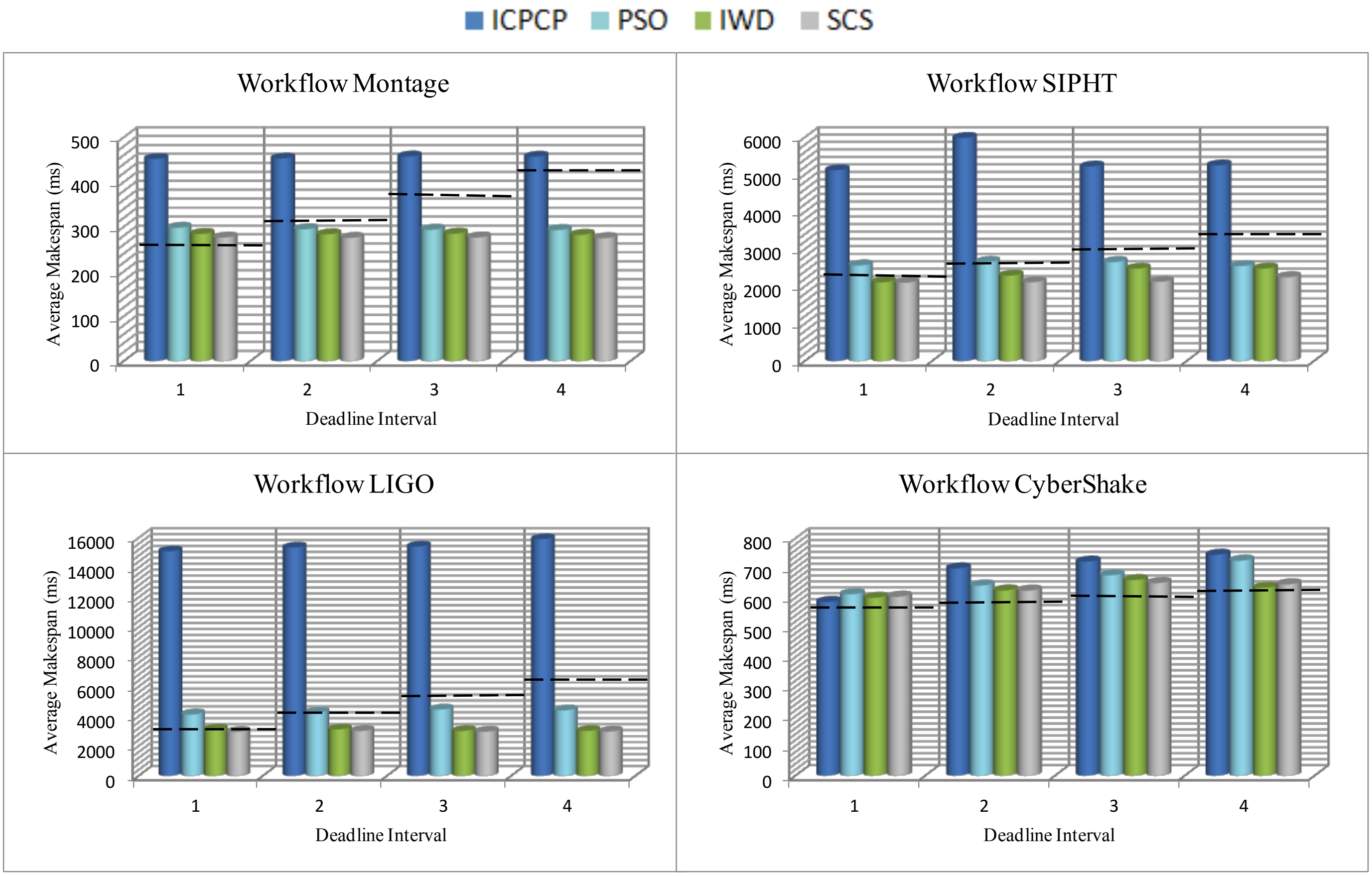

Fig. 5. Average Makespan of each workflow in each deadline interval. The dotted line shows the deadline in that deadline interval.

The IWD and SCS algorithm for Montage, LIGO and SIPHT workflow produces schedules which almost in 100 percent of cases meet the deadlines. For LIGO workflow our algorithm has outperformed all the three algorithms by producing the schedules with lowest average makespan.

For SIPHT application, IWD algorithm performs better than PSO algorithm by generating schedules with lower average makespan and there is very less difference in the average makespan of SCS algorithm and IWD algorithm. For CyberShake, the ICPCP algorithm performs comparatively better and produces schedules with average makespan less than the deadline in the first interval. Among the PSO, IWD and SCS algorithms for the first interval the IWD has the lowest average makespan. For the 2nd deadline interval the average makespan of IWD algorithm is very close to that of SCS algorithm and it is considerable less than ICPCP and PSO algorithms. The ICPCP algorithm again averages the highest for the interval 1 through 4. For the 4th interval again the IWD algorithm reports the lowest average makespan outperforming SCS, PSO and ICPCP algorithms.

In all, the analysis of the average makespan of the algorithms supports the results of analysis of the deadline constrained evaluation that ICPCP has defaulted the maximum number of workflow deadlines and is not efficient in meeting the workflow deadlines. In all the deadline intervals the IWD algorithm has outperformed the ICPCP and the metaheuristic PSO algorithm. Our algorithm produces schedules comparable to SCS algorithm and in most cases schedules are of smaller makespan as compared to SCS algorithm.

### 6.1.3 Cost Evaluation

Fig. 6 shows the average execution cost for each workflow and deadline interval. The Cost, Deadline graphs shows the average cost in the deadline interval and the corresponding graph below depicts the average makespan by the respective

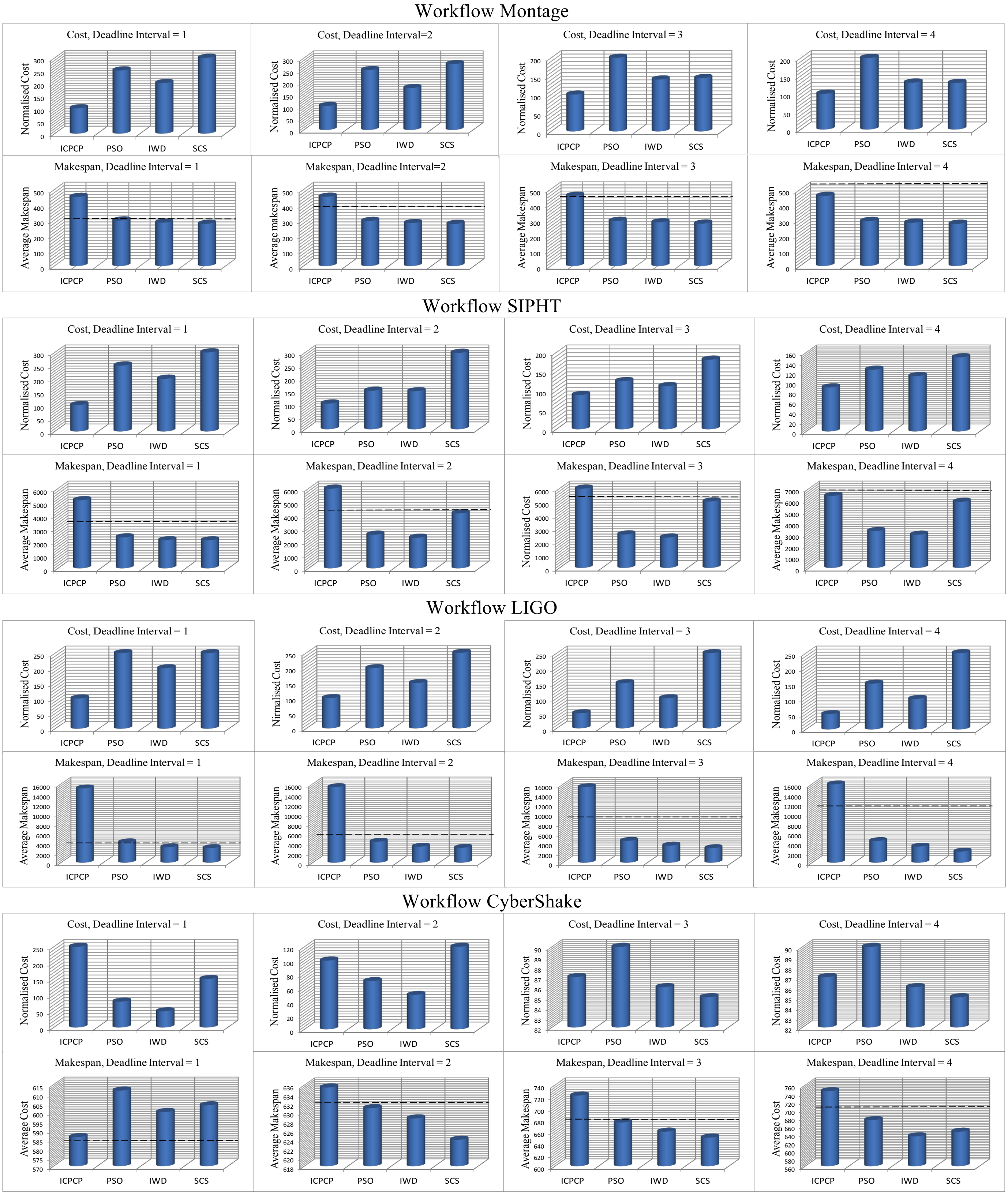

Fig. 6 Average makespan (ms) and normalised cost for each workflow and deadline interval. The dotted reference line in each panel depicts the deadline value of the respective deadline interval

algorithms. The dashed reference line marks the deadline for the workflow in the corresponding deadline interval. This cost versus makespan analysis helps us analyze that whether the algorithms are able to generate schedules within deadline with acceptable cost.

The ICPCP algorithm for Montage workflow generates the least cost schedules but it is not able to meet deadlines in any of the interval. The IWD algorithm generates least cost schedule in deadline interval 2 and 3 among all the algorithms and it also generates schedules with makespan much less than the deadline for these intervals. For the deadline Interval 4 our algorithm has generated least cost schedule with the least average makespan among all the algorithms. Although the average makespan for PSO, IWD and SCS are almost same but there is a pronounced difference between the average cost of schedules generated by these algorithms and for each of the deadline interval IWD has generated a schedule with least average cost.

For SIPHT workflow, our algorithm has performed the best by generating least cost schedules as compared to other three algorithms and meeting the deadlines of all the intervals with the least average makespan. Although the ICPCP algorithm has generated the schedules with cost less than IWD algorithm but the algorithm is unable to meet the deadline of any of the interval. The average makespan for PSO, IWD and SCS algorithm are all below the specified deadline for each of the interval so the algorithm which generates a schedule with least cost is preferable. IWD algorithm has not only generated schedules with least cost but there is a considerable difference in the cost of the schedules generated with PSO and SCS algorithm. So for the SIPHT application, IWD algorithm outperforms the ICPCP, PSO and SCS algorithm.

For LIGO workflow, all the algorithms follow the same trend as for the SIPHT workflow with ICPCP generating schedules with least cost but with makespan greater than the deadline for each of the interval. In this case also the performance of IWD algorithm is better than ICPCP, PSO and SCS algorithms as it schedules workflow tasks with least cost and least average makespan.

For CyberShake workflow, in the deadline interval 1 and 4 IWD algorithm has produced the most efficient schedule by producing schedules with least average cost and least average makespan. The ICPCP algorithm generates the schedules with very high cost as compared to SCS and IWD algorithm and its average makespan is also considerably high for all the deadline intervals. For the first deadline interval the PSO algorithm has generated the schedules with average makespan considerable higher than the deadline but it improves its performance over the next three deadline intervals. However the PSO algorithm generates highest cost schedules for the last two deadline intervals. From interval 2 through 4 the IWD and SCS algorithm has proved to be most consistent algorithms by generating least cost schedules.

In essence, it can be summarized form the above analysis that although ICPCP algorithm is able to generate the Cost effective schedules but it fails to schedule tasks within deadlines. In most of the cases the other three PSO, IWD and SCS algorithms efficiently schedules tasks but when compared among these algorithms IWD and SCS performs better by generating schedules with least cost within the deadline for each of the deadline interval.

#### 6.1.4 Cost-Deadline Tradeoff Analysis

In this part, we analyze the tradeoff between the resource provisioning cost and execution time of workflow when the workflows are executed in the IaaS cloud environment. We are interested in understanding the variance of cost when the deadline of the workflow application varies from most strict (deadline interval 1) to more liberal (deadline interval 4). For this analysis, we have plotted the cost of execution of the workflow application against the deadlines for each of the deadline interval as depicted in Figs. 7.

For Montage workflow, as depicted in Fig. 7(a) the ICPCP algorithm incurs constant costs for all the deadline intervals. For all the other three algorithms a converging trend towards minimizing costs is observed. The SCS, IWD and PSO algorithms approximately achieves 56, 35 and 25 percent cost reduction respectively. As is visible in Fig. 7(b) the same converging trend of cost reduction can be observed when deadline becomes less strict for SIPHT workflow for all the algorithms. For SIPHT, SCS and PSO algorithm almost halved the cost of execution when the deadline is relaxed from deadline interval 1 to deadline interval 4. IWD algorithm reduced the cost by 44 percent and ICPCP by 11 percent for the SIPHT workflow.

For LIGO workflow as Fig 7(c) shows, SCS algorithm do not reduces the cost of execution of workflow for relaxed deadlines over the first deadline interval to fourth deadline interval. In this case, PSO and IWD algorithm outperforms the ICPCP and SCS algorithm by saving almost 50 percent of cost for the end user. For the CyberShake Workflow the Fig. 7(d) shows that all the algorithms minimize costs as the deadlines become less strict.

The results of the experiment conducted to understand the cost-deadline tradeoff in cloud computing environments demonstrates that the algorithms schedule workflow tasks with less cost as the deadline are relaxed. This is possibly achieved by scheduling tasks on cheaper virtual machines which may also compromise the workflow deadlines.

This inherent tradeoff between QoS requirement satisfaction (workflow deadline in our case) and Resources provisioning cost optimization must be properly balanced to efficiently schedule scientific workflow on IaaS clouds.

#### 6.1.4 Further Discussion

##### 6.1.4.1 IWD Algorithm Complexity

The computational complexity of the IWD algorithm depends on the maximum number of iteration ($\mathrm{Max}_{\mathrm{Itr}}$)

decided for the algorithm to execute the number of VMs to visit $VM_{tovisit}$ and the number of tasks in the workflow which in our system model is the number of water drops (Iwd). Then for a particular iteration if these three factors are assumed to be n the complexity of the algorithm is of the order of $O(n^3)$.

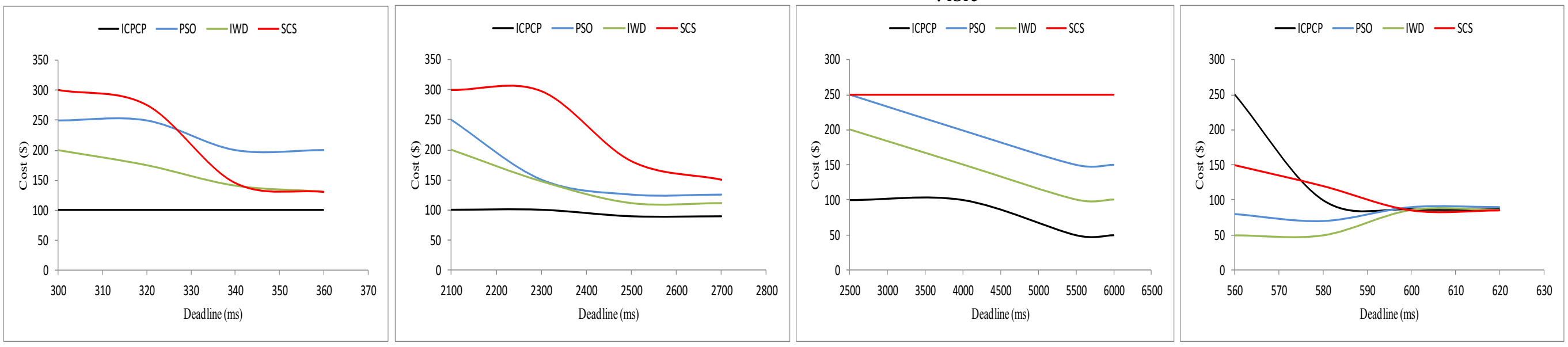

(a) (b) (c) (d)

Fig. 7. Cost-Deadline Tradeoff Analysis for workflows. (a) Montage. (b) SIPHT. (c) LIGO. (d) CyberShake

### 6.1.4.2 IWD Algorithm Makespan Convergence

The IWD algorithm converges towards optimal solution schedule as the number of iterations increases. Fig. 8 shows the convergence curve of the IWD algorithm when the number of iteration of the algorithm is plotted against the average makespan of each of the workflow over deadline interval 1. For each of the workflow the average makespan decreases monotonically as the number of iteration increases.

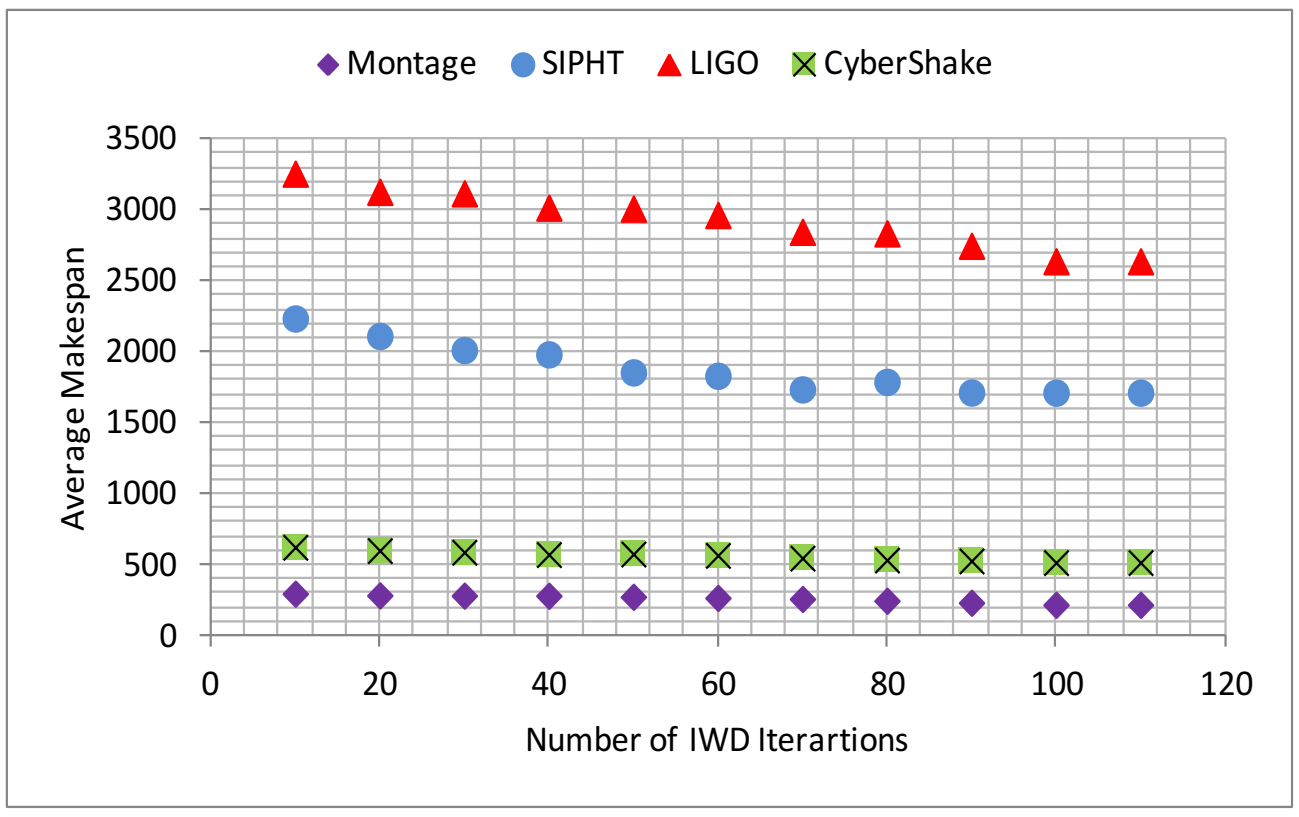

Fig. 8. Convergence Curve: Average Makespan of ten runs of IWD Algorithm

The results of convergence analysis of the algorithm show that the algorithm converges fast and needs less number of iterations to converge to good quality near optimal solution. Moreover, the algorithm in each iteration searches a fixed part of combinatorial search space which makes the algorithm computationally light.

## 7 Conclusion And Future Work

Scheduling large scale scientific workflow application on high performance computing environments like clouds is a fundamental NP-complete problem. However, most of the scheduling algorithms developed for large-scale distributed systems are best effort based. When complex data and compute intensive workflow application are executed on clouds the optimization of QoS parameters of cost and deadline becomes a crucial research challenge. In this research paper, we have studied the research challenge of provisioning resources to schedule the scientific workflow tasks within the deadlines with optimized cost in IaaS cloud environment. For this purpose we have modeled the resource provisioning and scheduling problem as a joint optimization problem and have proposed a resource provisioning and scheduling algorithm with the help of Intelligent Water Drop algorithm. The main contribution of our research is the development of a resource provisioning and scheduling framework with the help of which scientists will be able to exploit the capabilities of cloud computing in a cost efficient manner.

We have evaluate IWD algorithm with four real time workflow applications and experimental results shows that IWD algorithm in most strict deadlines meets 97% percent of deadlines for all the workflows and outperforms the ICPCP and PSO algorithms. In the relaxed deadline intervals of 2, 3 and 4 IWD algorithm achieves maximum performance by meeting 100% of deadlines for all the workflows.

A solution whose makespan is below the deadline is favored as compared to the solution which overshoots the deadline. The IWD algorithm in the most strict deadline interval 1 achieves makespan less than deadline in more than 96% of times better than ICPCP and PSO algorithm and close to SCS algorithm for all workflows. In all the other deadline intervals IWD algorithm schedules tasks with lower makespan in more than 95% of cases. Experimental results for cost evaluation of ICPCP, PSO, SCS and our algorithm for Montage, LIGO and SIPHT workflows demonstrates that IWD algorithm on an average achieves 76% lower monetary cost for all the deadline intervals.

When the average makespan is plotted against the increasing number of iterations of the algorithm it was found that IWD algorithm converges towards optimal solution by achieving lower average makespan with the increasing number of iterations. As future work, we will tune the various static and dynamic parameters of the IWD algorithm to achieve the enhanced performance in even less number of iterations. Another direction which can be further explored is the application of different initial resource selection strategies.